\documentclass[a4,useAMS,usenatbib,usegraphicx]{mn2e}
\def\aapr{\ref@jnl{A\&A~Rev.}}		

\usepackage{amsmath}
\include{epsf}
\usepackage{float}
\usepackage{color}

\usepackage{epsf}
\epsfclipon
\title[Surviving infant mortality]{Surviving infant mortality in the hierarchical merging scenario}
\author[R.Smith et al]{R.Smith$^{1}$\thanks{E-mail:rsmith@astro-udec.cl}, M. Fellhauer${^1}$, S. Goodwin${^2}$, P. Assmann${^1}$\\
$^{1}$Departamento de Astronomia, Universidad de Concepcion, Casilla 160-C, Concepcion, Chile\\
\noindent
$^{2}$Department of Physics and Astronomy, University of Sheffield, Hicks Building, Hounsfield Road, Sheffield, S3 7RH, UK}
\begin{document}

\date{Accepted 2011 February 25. Received -----; in original form -----}

\pagerange{\pageref{firstpage}--\pageref{lastpage}} \pubyear{2011}

\maketitle

\label{firstpage}

\begin{abstract}
We examine the effects of gas expulsion on initially sub-structured
and out-of-equilibrium star clusters.  We perform $N$-body simulations
of the evolution of star clusters in a static background potential
before removing that potential to model gas expulsion.  We find that
the initial star formation efficiency is not a good measure of the
survivability of star clusters.  This is because the stellar
distribution can change significantly, causing a large change in the 
relative importance of the stellar and gas potentials.  We find that
the initial stellar distribution and velocity dispersion are far more
important parameters than the initial star formation efficiency, and
that clusters with very low star formation efficiencies can survive
gas expulsion.  We suggest that it is variations in cluster initial
conditions rather than in their star formation efficiencies that cause
some clusters to be destroyed while a few survive.
\end{abstract}


\section{Introduction}

Many, perhaps the vast majority, of stars form in dense, clustered
environments (\citealp{Lada2003}; \citealp{Bressert2010}).  However,
after only 10 -- 20~Myr the vast majority of stars are dispersed into 
the low density environment of the field (\citealp{Lada2003}).  The
destruction of most of these early dense clusters has been termed
`infant mortality'.

Perhaps the best candidate for the mechanism behind infant mortality
is primordial gas expulsion.  Giant molecular clouds turn only a few
per cent to a few tens of per cent of their gas into stars (\citealp{Lada2003})
meaning that the potentials of very young embedded clusters are
dominated by gas.  Feedback from the most massive stars will remove
this gas on a timescale of a few~Myr significantly altering the
potential of the cluster and potentially leading to its destruction.
This process has been studied analytically and with simulations 
by many authors (e.g. \citealp{Tutukov1978}; 
\citealp{Hills1980}; \citealp{Mathieu1983}; \citealp{Elmegreen1983}; \citealp{Lada1984}; 
\citealp{Elmegreen1985}; \citealp{Pinto1987}; \citealp{Verschueren1989};
\citealp{Goodwin1997a}; \citealp{Goodwin1997b}; \citealp{Geyer2001}; \citealp{Boily2003a}; \citealp{Boily2003b}; \citealp{Bastian2006}; \citealp{Goodwin2006}; \citealp{Baumgardt2007}; \citealp{Parmentier2008}; \citealp{Goodwin2009ApSS}; \citealp{Chen2009}).

However, previous work has tended to concentrate on clusters that are
both in virial equilibrium (notable exceptions are \citealp{Lada1984}; 
\citealp{Elmegreen1985}; \citealp{Verschueren1989}; \citealp{Goodwin2009ApSS})
and structurally simple (usually a Plummer sphere or similar).  But
recent observational and theoretical results strongly suggest that
stars do not form in dynamical equilibrium, nor in a smooth
distribution (see e.g. \citealp{Elmegreen2001}; \citealp{Allen2007}; 
\citealp{Gutermuth2009}; \citealp{Bressert2010} and references in all of
these papers).  This should not be surprising as in the gravoturbulent
model of star formation, stars will form in dense gas in filaments and
clumps in a turbulent environment (see e.g. \citealp{Elmegreen2004}).  This
has lead to a model of star cluster formation in which a dynamically
cool and clumpy initial state collapses and violently relaxes into a
dense star cluster (e.g. \citealp{Allison2009b}, \citealp{Allison2010}).

Strongly non-equilibrium initial conditions may play a crucial role in
the effects of gas expulsion on clusters because the key factor in
determining the effects of gas expulsion is not only the depth of the
gas potential, but also the dynamical state of the stars {\em 
at the onset of gas expulsion} (\citealp{Verschueren1989}; \citealp{Goodwin2009ApSS}).  As an extreme example, if 99 per cent of the mass of a cluster
is expelled, the cluster will still survive if the stars have almost
no velocity dispersion.

We might therefore expect the effects of gas expulsion on a cluster to
depend on three factors.  Firstly, a clumpy and non-symmetrical initial distribution of the stars. Secondly, the star formation efficiency (SFE), and
so the relative importance of the stellar and gas potentials.
Thridly, the (non-equilibrium) initial conditions of the cluster and
the evolution of these initial conditions up until the point of gas
expulsion.  If clusters are unable to relax fully before gas expulsion
then we might expect them to respond very differently to the loss of
the residual gas.

In this paper we examine the $N$-body evolution of highly non-equilibrium 
star cluster in fixed background potentials to model the background gas.
The background potential is then removed to simulate the effects of
gas expulsion.  We describe our initial conditions in Section~2, our
results in Section~3, and then discuss the potential consequences in
Section~4 before drawing our conclusions in Section~5.

\section{Initial conditions}

We perform our $N$-body simulations using the Nbody2 code
(\citealp{Aarseth2001}). Nbody2 is a fast and accurate
direct-integration code, optimised for the number of star particles
used in this study ($N = 1000$).  In our simulations there are 
two separate mass distributions which we wish to model: the
stars and the background gas potential.

\subsection{Stellar distribution}

In all cases we model the stellar distribution as $N=1000$ particles
with equal masses of $0.5 M_\odot$ resulting in a total stellar mass
of $500 M_\odot$.  Every particle has a gravitational softening length
of 100~AU.  We choose softened and equal-mass particles in order to
avoid strong two-body interactions and mass segregation. \cite{Allison2009b} and \cite{Allison2010} showed that both of these effects can be extremely
important in the violent collapse of cool, clumpy regions, however we
wish to avoid complicating our simulations with these effects as we
are interested in the effects of gas expulsion.

\begin{figure*}
\begin{center}$
\begin{array}{cc}
\includegraphics[width=3.2in]{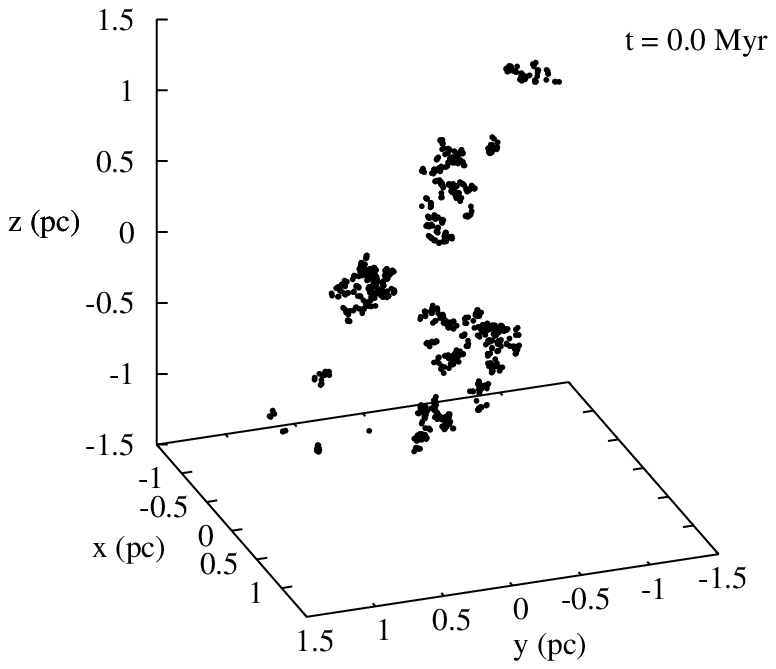} &
\includegraphics[width=3.2in]{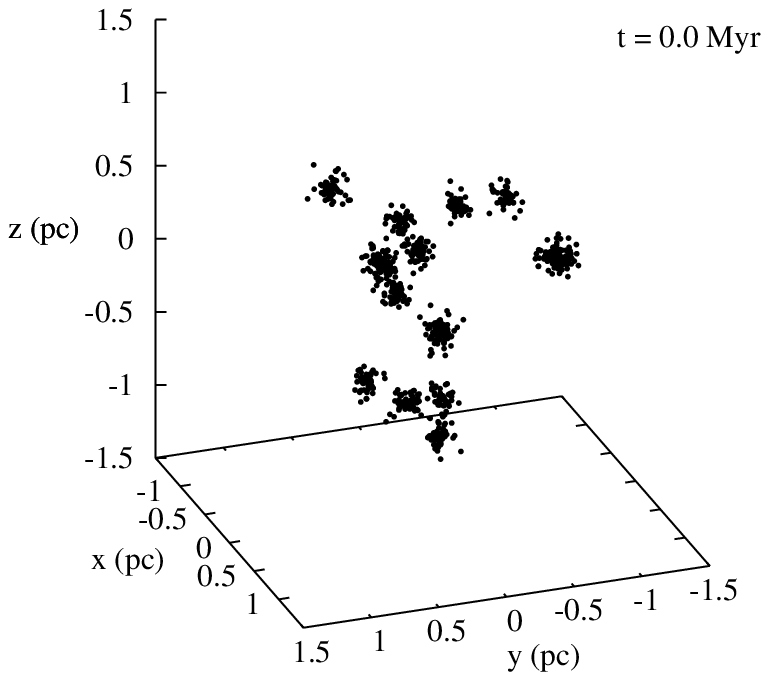} \\
\end{array}$
\end{center}
\caption{Representative examples of an initial distribution of stars
  with a fractal morphology (left panel), and a clumpy plummer morphology
  (right panel).}
\label{morphrep}
\end{figure*}

We distribute the stars within a radius of 1.5~pc with two different
`clumpy' morphologies: `fractal', and `clumpy plummer'.
Representative snapshots of the initial
conditions of a fractal and clumpy plummer distribution can be seen in the
left and right-hand panels, respectively, of Fig.~\ref{morphrep}.

Fractal clusters are produced using the box fractal technique
described by \cite{Goodwin2004}; also used by \cite{Allison2009a} and \cite{Allison2010}.  In this paper we use a fractal dimension $D = 1.6$
which corresponds to a highly clumpy initial distribution (see 
Fig.~\ref{morphrep}).

In clumpy plummers, each sub-clump is modelled as an individual
Plummer sphere. The total stellar mass of 500 M$_\odot$ is distributed
into 16 sub-clumps of approximately equal mass. Each sub-clump
contains 31.0-31.5 M$_\odot$ formed from 62-63 
particles. The position of individual sub-clumps within the model star
forming region follows the gas potential (see below). 

\subsection{Gas distribution}

The gas within the star-forming region is modelled as a static Plummer
potential with a mass $M_{\rm{g}}$ and scale-radius $r_{\rm{g}}$.

The scale radius is set to be $r_{\rm{g}} = 1$~pc, or $r_{\rm{g}} = 1.5$~pc.  As
the stars have a maximum radius of 1.5~pc this means that for $r_{\rm{g}} =
1.5$~pc, the gas distribution seen by the stars is roughly uniform. 

The relative masses of the gas potential within 1.5~pc and 
the stars set the {\em true} SFE ($\epsilon$), that is the fraction of 
the initial cloud mass converted into stars.  The mass of the gas 
potential is varied to obtain true SFEs of 20,
30 and 40 per cent ($M_{\rm{g}}=$2000, 1167, 750~$M_\odot$).  

Note that a true SFE of 20 or 30 per cent will
fail to produce a bound star cluster after instantaneous gas expulsion if the gas
and stars are initially virialised (see e.g. \citealp{Baumgardt2007}).  

We emphasise that the gas potential does not follow the initial stellar
distribution, nor does it react to changes in the stellar distribution
as it evolves (it is not live).  These are obviously extreme
simplifications, but as we will discuss later we feel that we capture
the essence of the basic physics using such a simple model.

\subsection{The dynamical state of the stars}

The virial ratio $Q = T/|\Omega|$ is the ratio of the kinetic , $T$, to
potential energy, $\Omega$, of the system.  We set the initial
velocity dispersion of the stars relative to the total
potential (gas \& stars) to set initial stellar virial ratios between zero (cold),
0.5 (virial equilibrium), and 0.9 (supervirial, but
bound).  

For initial virial ratios $Q<0.5$, the system will tend to 
collapse, and for $Q>0.5$ the system will tend to expand.  However,
even for $Q=0.5$, although the system is in virial equilibrium, 
the clumpy initial conditions mean that it is {\em not} in dynamical 
equilibrium.

\subsection{Gas expulsion}
Although young star clusters form embedded within the the molecular gas from which they formed, few star clusters over $\sim 5$~Myr old remain associated with their gas (\citealp{Proszkow2009}). This is likely as a result of a number of mechanisms including radiative feedback from massive stars, stellar winds from young stars, and eventually the onset of the first supernova. The time at which gas removal begins to occur, and the duration of the gas removal process is uncertain, and dependent on the particular gas removal mechanism in operation.

To simulate gas expulsion we instantaneously remove the external gas
potential after 3~Myr. This is approximately mid-way between the time at which gas removal from stellar winds and supernova feedback might be expected to occur for star clusters containing $\sim1000$ stars. Instantaneous gas removal is the most extreme
form of gas removal as the stars have no chance to readjust to the
change in the potential (e.g. \citealp{Goodwin1997a}; \citealp{Baumgardt2007}).

\subsection{Ensembles}

\begin{figure*}
\begin{center}$
\begin{array}{cc}
\includegraphics[height=8cm,width=9cm]{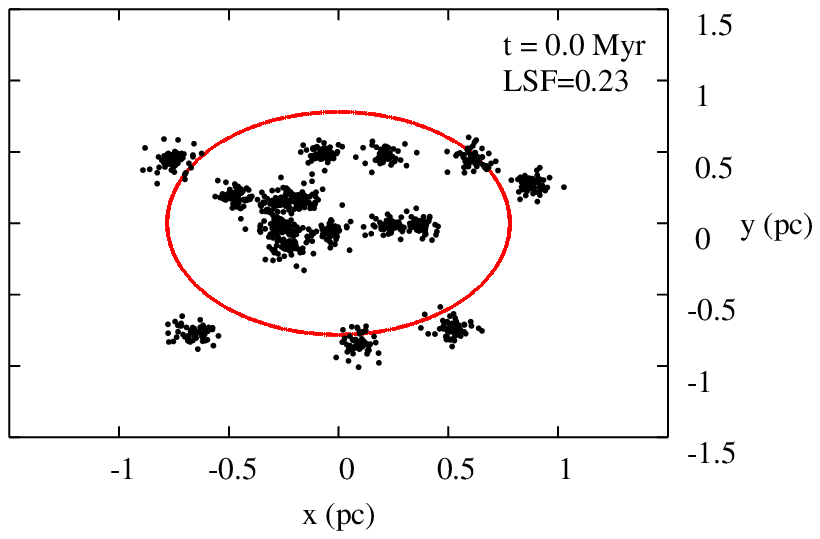} &
\includegraphics[height=8cm,width=9cm]{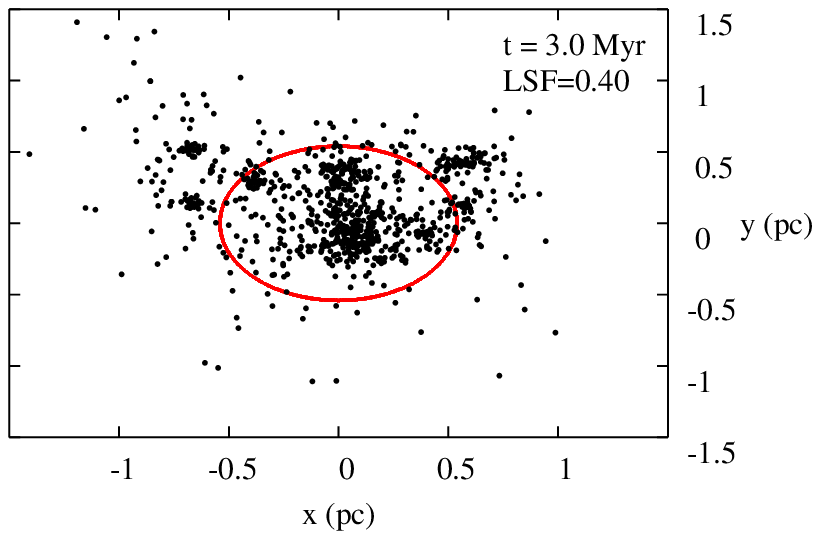} \\
\end{array}$
\end{center}
\caption{An x-y projection of the initial (left panel) and final (right panel) stellar distribution of an initially clumpy plummer morphology stellar
  distribution, evolved for 3 Myr in an $r_{\rm{g}} = 1$~pc, $M_{\rm{g}} = 2000 M_\odot$ gas
  potential. The initial virial ratio of the stellar distribution is 0.35. Time of snapshot and LSF at that instant is indicated in top-right hand corner of
  each panel. The dotted (red) circle marks the stellar half-mass
  radius, directly used to calculate the $LSF$. In this case, the
  $LSF$ almost doubles from 0.23 (left panel) to 0.40 (right panel) over the
  duration of the embedded phase.}
\label{plum_rh}
\end{figure*}

It should be noted that both the clumpy plummer and fractal clusters 
can vary considerably in
appearance depending on the random realisation used and the subsequent
evolution is highly stochastic (see \citealp{Allison2010}). We therefore
conduct a minimum of 5 random realisations of each parameter set.

\subsection{Summary}

We set-up clumpy $500 M_\odot$ star clusters with $N=1000$ equal-mass
particles within a static background gas potential.  The dynamical
state of the stars varies from very cold to almost unbound.  The stars
dynamically evolve within the gas potential before its instantaneous
removal after 3~Myr. We measure the final properties of each star cluster an additional 3~Myr after gas expulsion. A summary of the key
parameters is provided in Table \ref{parstable}.

\begin{table}
\centering

\begin{tabular}{|c|c|}
\hline Total mass, $M_{\rm{tot}}$ & 1250-2500 M$_\odot$* \\ Outer radius &
1.5 pc \\ Star formation efficiency, $\epsilon$ & 20-40$\%$* \\ \hline
Total stellar mass, $M_{\rm{star}}$ & 500 M$_\odot$ \\ Particle
number, $N$ & 1000 \\ Particle mass, $m_{\rm{part}}$ & 0.5
M$_\odot$ \\ Stellar morphology & fractal or Plummer *\\ Initial
virial ratio, $Q_{\rm{i}}$ & 0.0-1.0* \\ \hline Total gas mass, $M_{\rm{g}}$ & 750-2000 M$_\odot$* \\ Gas Plummer scale radius,
$r_{\rm{g}}$ & 1.0 pc or 1.5 pc*\\ crossing-time,
$\tau_{\rm{cr}}$ & 1.4 Myr/2.0 Myr \\ &($r_{\rm{g}}$=1.0
pc/1.5 pc)\\ \hline
\end{tabular}
\caption{The initial conditions parameter sets. Parameters are loosely
  divided into quantities governing; the total region (upper),the
  stellar component (middle), the gas component (lower). Parameters
  that are highlighted with a `*' are parameters we adjust between
  simulations.}
\label{parstable}
\end{table}

\section{Results}

We will first discuss the evolution and relaxation of the stellar
distribution in the gas potential before addressing the response of
the stars to gas expulsion.

\subsection{The embedded phase}
\begin{figure*}
\begin{center}$
\begin{array}{cc}
\includegraphics[height=7cm,width=9cm]{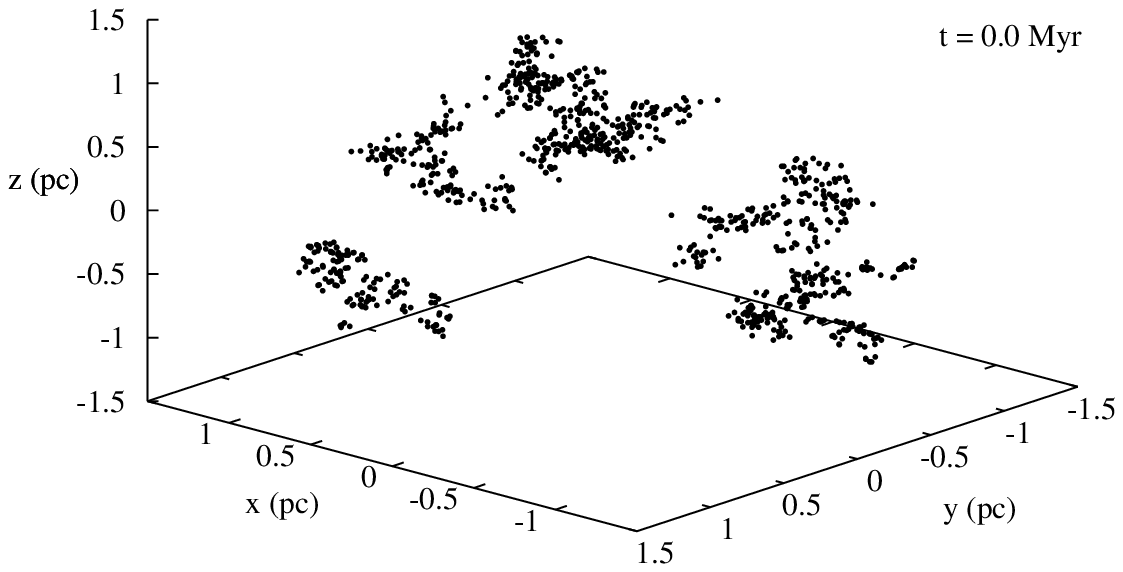} &  
\includegraphics[height=7cm,width=9cm]{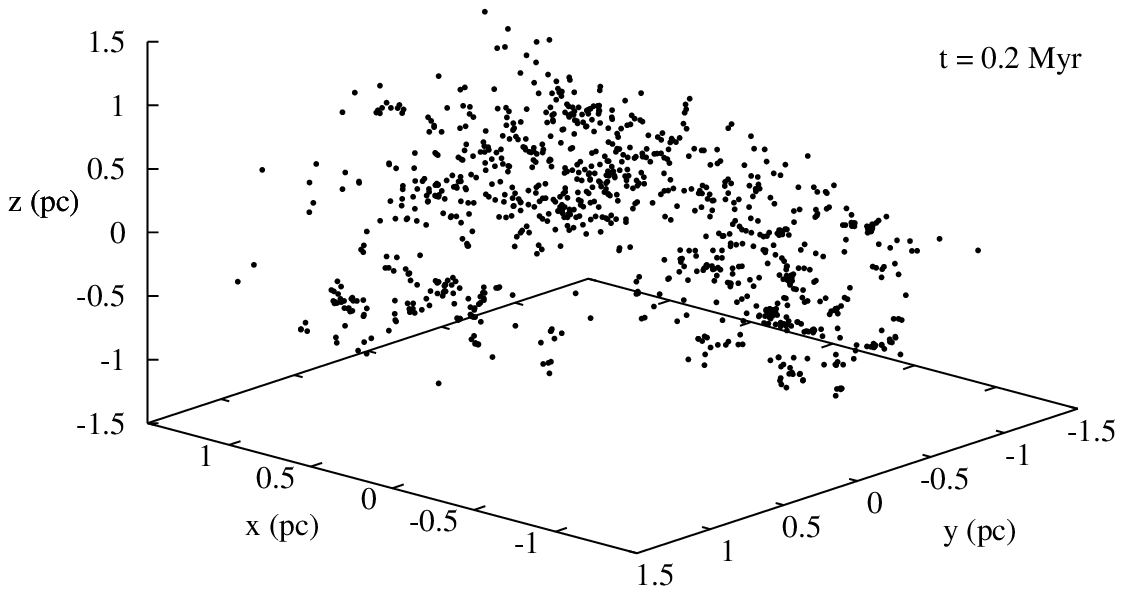} \\ 
\includegraphics[height=7cm,width=9cm]{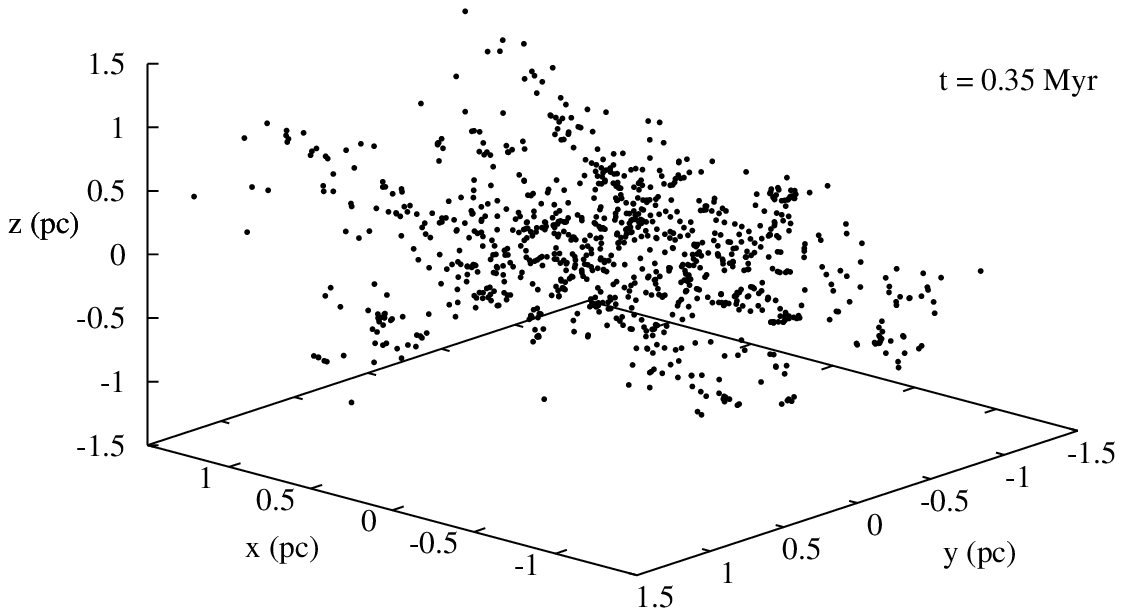} &  
\includegraphics[height=7cm,width=9cm]{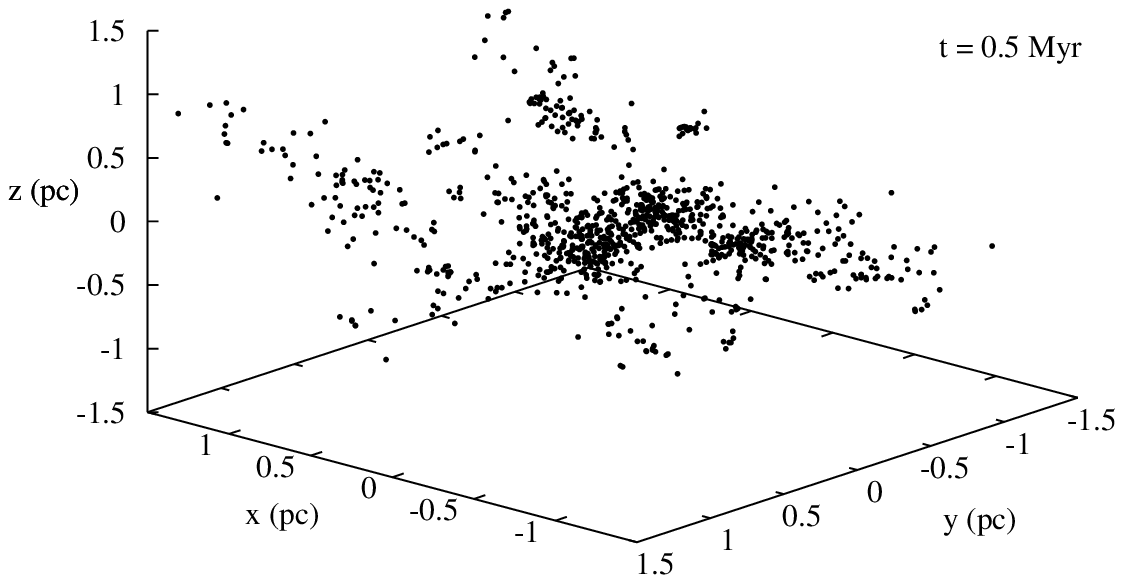} \\
\includegraphics[height=7cm,width=9cm]{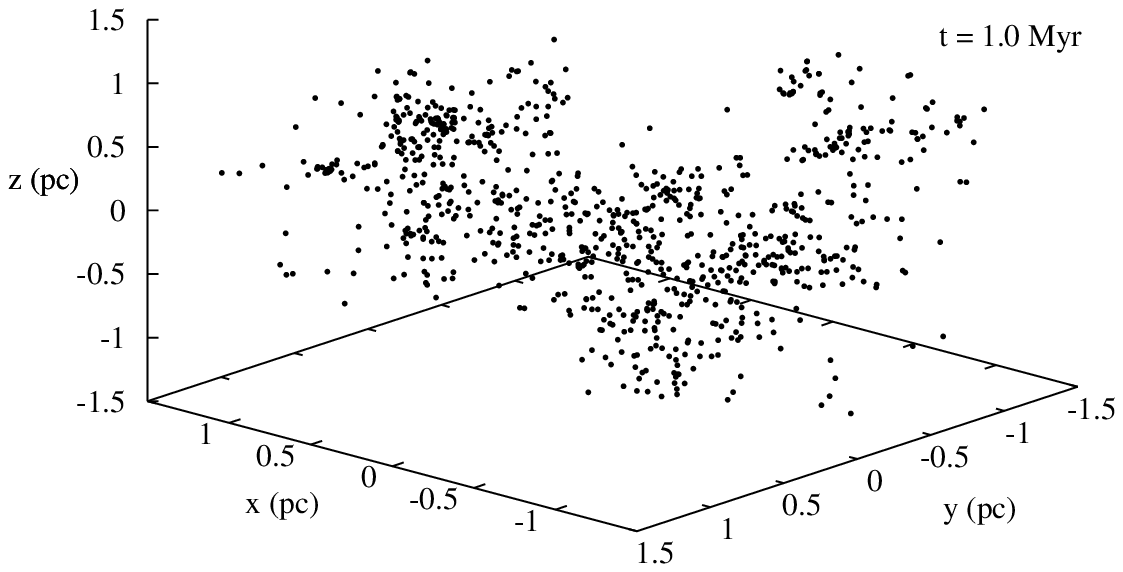} &
\includegraphics[height=7cm,width=9cm]{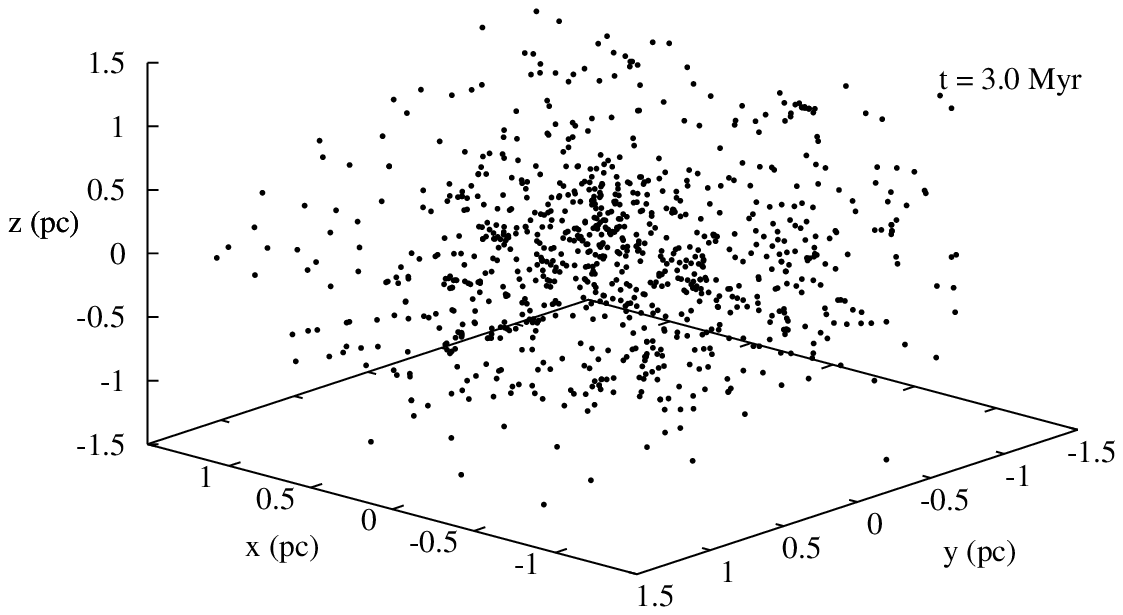} \\
\end{array}$
\end{center}
\caption{The evolution of an initially fractal ($D=1.6$) stellar
  distribution in an $r_{\rm{g}} = 1$~pc, $M_{\rm{g}} = 2000 M_\odot$ gas
  potential.  The initial virial ratio of the stellar distribution is
  0.33. The time of each snapshot is indicated in top-right hand corner of each panel.}
\label{reprun}
\end{figure*}

In all simulations, we evolve our stellar initial conditions for 3 Myr
in the gas potential.  In Figure \ref{reprun} we present an example 
of the evolution of an initially fractal stellar distribution with an
initial virial ratio of 0.33 in a gas potential of scale-radius 1~pc. The initial true SFE of this system is $\epsilon=20\%$. 

Clumpy stellar sub-structure interacts resulting in clump merging, and
stellar scattering from the clumps (somewhat suppressed in these
simulations due to the softening). As a result, substructure is significantly erased on
$\sim$1-2 crossing-times of the region. For example, the
crossing-time of the region presented in Fig.~\ref{reprun} is 1.4
Myr. By 3 Myr (bottom-right panel) over 2 crossing-times
have elapsed, and the remaining stellar distribution has lost most of its 
sub-structure. Similar behaviour is seen in the
simulations of \cite{Goodwin2004}, \cite{Allison2009b} and \cite{Allison2010}. However, they do not include a
model for the gas component of the region in their simulations. By
including a gas potential in our simulations, the crossing-time of the
region is shortened. However the picture of evolution from clumpy
substructure to an almost smooth distribution of stars, on timescales of
$\sim$1-2 crossing-times of the region, is unsurprisingly 
unchanged by the presence of the gas.

Importantly, the relaxation and smoothing of the stellar distribution
has resulting in it shrinking significantly (see also \citealp{Allison2009b}; \citealp{Allison2010}).  It is clear in Fig.~ \ref{reprun} that the central regions of the cluster now contain a greater number of stars than initially. This means that (because the gas potential is static) that the stars
are a more significant contributor to the central potential than
initially - the {\em effective SFE} has
increased (\citealp{Verschueren1989}; \citealp{Goodwin2009ApSS}) and, in this case,
the cluster should be more able to survive gas expulsion.

In order to quantify the effects of collapse (or expansion) and the
change of the effective SFE we define 
the {\it{ Local Stellar Fraction}} ($LSF$) of the star forming
region:
\begin{equation}
\label{LSFeqn}
LSF=\frac{M_{\star}(r<r_{\rm{h}})}{M_{\star}(r<r_{\rm{h}})+M_{\rm{gas}}(r<r_{\rm{h}})}.
\end{equation}
where $r_{\rm{h}}$ is the radius from the centre of the region
containing half the total mass of {\em stars}. $M_{\star}$
and $M_{\rm{gas}}$ is the mass of stars and gas, respectively,
measured within $r_{\rm{h}}$. Therefore the LSF is a measure of the
current effective SFE.

In any non-equilibrium system the half-mass radius of the stars will
change with time.  Figure~\ref{plum_rh} shows a projection of the
stellar distributions of an initially clumpy plummer model within an $r_{\rm{g}} = 1$~pc, $M_{\rm{g}} = 2000 M_\odot$ gas potential. Initially the stars have an initial virial ratio of 0.35, and a half-mass radius $r_{\rm{h}}=0.78$~pc. $LSF = 0.23$, close to the initial 
true SFE of $\epsilon = 0.2$ (left panel). However, at 3~Myr the stellar half-mass radius has fallen to $r_{\rm{h}} \sim 0.55$~pc and the $LSF$ is $0.73$ (right panel) - almost double the initial value.

\begin{figure}
  \centering \epsfxsize=8.5cm \epsfysize=7cm \epsffile{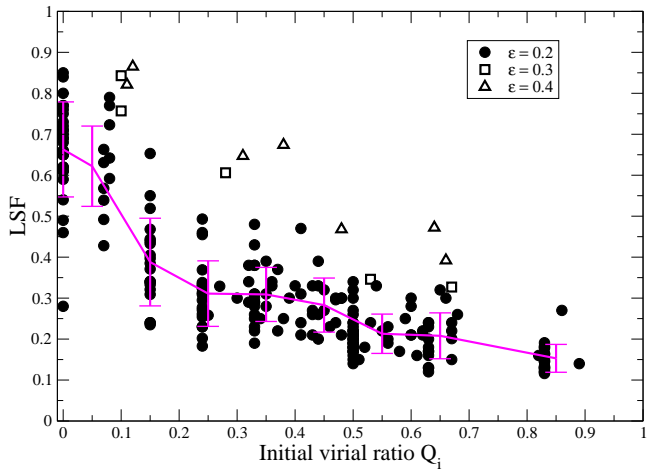}
  \caption{Plot of the initial virial ratio $Q_{\rm{i}}$ against the Local Stellar 
    Fraction ($LSF$) at 3~Myr. Symbols reflect the initial true SFE 
    of the region of 20, 30 or 40 per cent. Initially hot
    dynamical temperatures result in low $LSF$, and conversely very
    cold dynamical temperatures result in high $LSF$. At a fixed value
    of initial virial ratio, there is a slight trend for raised $LSF$
    for increasing true SFEs $\epsilon$. Error bars show the mean and standard deviation of the scatter observed in the simulations with an initial true SFE of 20 per cent.}
\label{QiLSF}
\end{figure}

\begin{figure}
  \centering \epsfxsize=8.5cm \epsfysize=7cm \epsffile{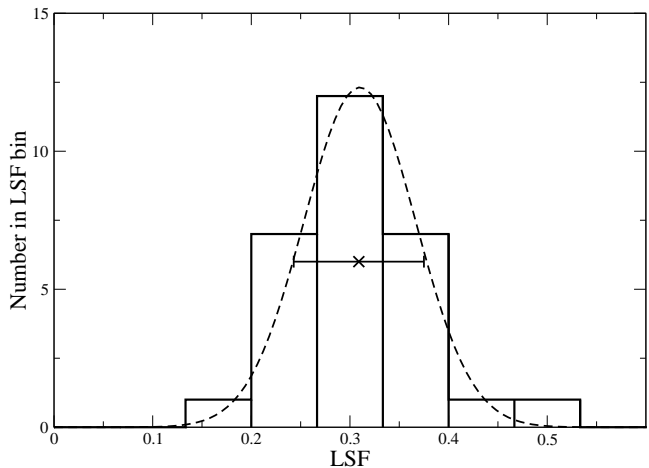}
  \caption{A representative histogram of the scatter in LSF within $Q_{\rm{i}}$ bin 0.1-0.2. Solid bars represent the numbers counted within each LSF bin. The cross point is the average LSF for all clusters within the bin. The error bars of the crossed point are the standard deviation of the LSF scatter. The dashed line shows a gaussian fit with equal standard deviation.}
\label{histLSF}
\end{figure}

In Figure \ref{QiLSF} we plot the $LSF$ measured moments
before gas expulsion, versus the initial virial ratio of the stars
$Q_{\rm{i}}$. Unsurprisingly, stellar distributions with low initial virial
ratio form clusters with high $LSF$s.  What is slightly surprising is
that high-$Q_{\rm{i}}$ (warm) clusters do not have significantly lower $LSF$s
than their initial true SFE.  This is because
the gas potential dominates and even almost unbound clusters only
expand by a factor of less than two in their half-mass radii (some
stars are lost, but the half-mass radii do not change by very
significant factors).  However, there is considerable 
scatter in the figure. The values with errorbars connected by a solid line in Figure \ref{QiLSF} represent the mean values with standard deviation of this scatter in LSF (for $\epsilon=0.2$ simulations) measured in 0.1 wide bins of $Q_{\rm{i}}$. We show a representative histogram for the $Q_{\rm{i}}$=0.3-0.4 bin in Figure \ref{histLSF}. On top of our 5 random realisations of each point of our parameter space, we conduct an additional 60 Plummer and 70 fractal morphology simulations so as to ensure a minimum of 20 simulations in each $Q_{\rm{i}}$ bin to avoid low number statistics. The scatter is higher for stellar distributions with low initial virial
ratios. This scatter reflects the stochastic nature of
cluster evolution from clumpy initial conditions (see \citealp{Allison2010}).  

We find no evidence for any dependency in the position or scatter of
the trend on initial stellar morphology (clumpy plummer or fractal) or gas
concentration (uniform or concentrated).  There is a trend in that
clusters with higher true SFEs have higher
$LSF$s reflecting their initially higher SFEs (open boxes and triangles in Figure \ref{QiLSF}).

As well as the $LSF$, the virial ratio of the stars 
at the onset of gas expulsion is also important.  In 
particular, the velocity dispersion
of the stars is crucial, with sub-virial clusters much more able to
survive gas expulsion (\citealp{Goodwin2009ApSS}).  Therefore we measure the
stellar virial ratio just before gas expulsion $Q_{\rm{pe}}$.

\begin{figure}
  \centering \epsfxsize=8.5cm \epsfysize=7cm \epsffile{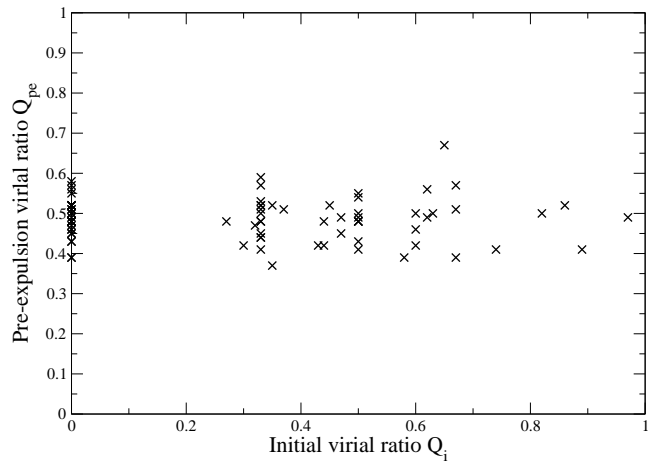}
  \caption{Comparison of the initial virial ratio $Q_{\rm{i}}$, chosen
    at the beginning of the simulation, versus the pre-gas-expulsion
    virial ratio $Q_{\rm{pe}}$, measured just prior to gas
    expulsion. Pre-expulsion virial ratios are all close to virialised
    ($Q_{\rm{pe}}=0.5$) but scatter around this value with a standard
    deviation $\sim0.05$.}
\label{QfQi}
\end{figure}

Figure~\ref{QfQi} shows the initial and pre-expulsion virial ratios
for all our clusters.  As would be expected, our clusters have
attempted to reach equilibrium (by erasing substructure and
collapsing/expanding).  In all cases the clusters have reached an {\em
  approximate} virial equilibrium of $Q_{\rm{pe}} \sim 0.5$.
However, it is crucial to note that these are approximate
equilibriums, and none of these clusters are fully virialised. We shall return to
this point later.

%%%%%%%%%%%%%%%%%%%%%%%%%%%%%%%%%%%%%%%%%%%%%%%%%%%%%%%%%%%%%%%%%%%%%%
\subsection{Post-gas-expulsion evolution}
After 3 Myr of evolution within the embedded phase (i.e. within the
gas background potential), we assume the gas is instantaneously 
expelled from the star
forming region. This is simply modelled by an instantaneous removal of
the gas potential. We further evolve the gas-free star cluster an
additional 3 Myr after gas expulsion. At this time, we measure the
bound mass of the star cluster. Any star whose kinetic energy is less
than it's potential energy is considered bound. We normalise the bound
mass by the total stellar mass in the simulation - this quantity is
referred to as the {\it{bound stellar fraction}} ($f_{\rm{bound}}$).

\begin{figure}
  \centering \epsfxsize=8.5cm \epsfysize=7cm \epsffile{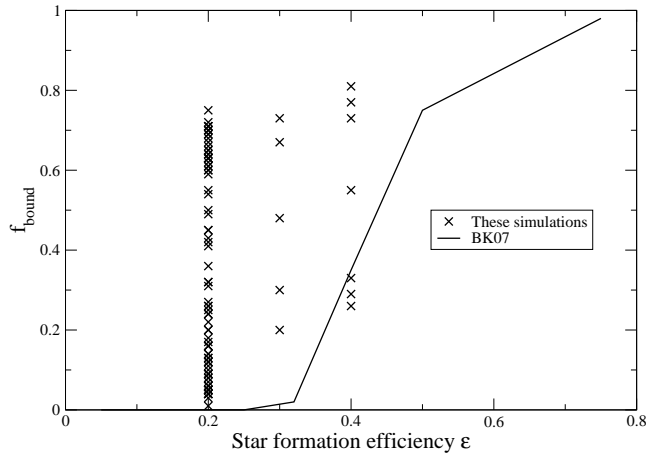}
  \caption{Plot of the region's SFE $\epsilon$
    versus the final cluster's bound mass fraction
    $f_{\rm{bound}}$. Crossed symbols are the results of all simulations in
    this paper, where as the solid line is the results of
    the BK07 simulations. The points are
    highly scattered with no clear trend, and do not match the results
    seen in the BK07 simulations.}
\label{fboundsfe}
\end{figure}

In Figure \ref{fboundsfe} we plot the bound stellar fraction
$f_{\rm{bound}}$, as a function of the {\em initial} true 
SFE $\epsilon$. We also plot the results of
\cite{Baumgardt2007} (herein BK07) who use initial smooth and
virialised equilibrium clusters as their initial conditions. 

It is quite clear from this figure that our clusters with an initial
true SFE of 20 or 30 per cent can survive gas expulsion
with very significant fractions of their initial mass remaining.  Also
there is a very significant scatter in our results rather than the
very tight relationship found by BK07.

We confirm that our choice of 3 Myr of gas free evolution has not significantly influenced our measured bound fraction by evolving a small sub-sample of clusters for 15 Myr after instantaneous gas removal. We continue the gas free evolution phase of three clusters that result in a high ($f_{\rm{bound}}=0.72)$), medium ($f_{\rm{bound}}=0.45)$) and low mass cluster ($f_{\rm{bound}}=0.30)$) when measured after 3 Myr. After 15 Myr, the measured bound fraction differs by $<5\%$ from those measured after 3 Myr.   

\begin{figure}
  \centering \epsfxsize=8.5cm \epsfysize=7cm \epsffile{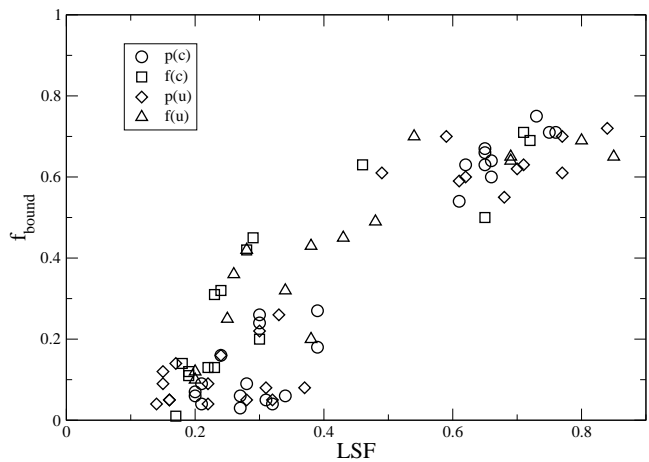}
  \caption{Plot of the influence of $LSF$ (measured at the onset of
    gas expulsion) on the final bound mass fraction ($f_{\rm{bound}}$)
    of clusters. All open symbols represent standard model simulations
    with varying initial stellar morphology or gas potential
    concentration. Key labels are fractal (f) or clumpy plummer (p)
    morphology, and uniform (u) or concentrated (c) gas
    concentration. There is a good trend for increasing bound mass
    fraction with increasing LSF, although clearly there is scatter in
    the trend. The trend is independent of initial stellar morphology
    or gas concentration}
\label{mnormLSF}
\end{figure}

The differences in our results from those of BK07 are in fact due to the differences in
the initial conditions used for the simulations.  As we have seen in
the previous section, the effective SFE as
measured by the $LSF$ can increase significantly before gas
expulsion.  Rather than expecting a relationship with the initial SFE 
we should expect a relationship with the $LSF$
which we plot in Fig.~\ref{mnormLSF}.  Here we find that the final bound
fraction of stars does correlate with the $LSF$ in a similar way to
the results of BK07.  Note that there is no systematic differences
between the results of simulations of fractals or clumpy plummers, nor
with the concentration of the gas potential.

\begin{figure}
  \centering \epsfxsize=8.5cm \epsfysize=7cm
  \epsffile{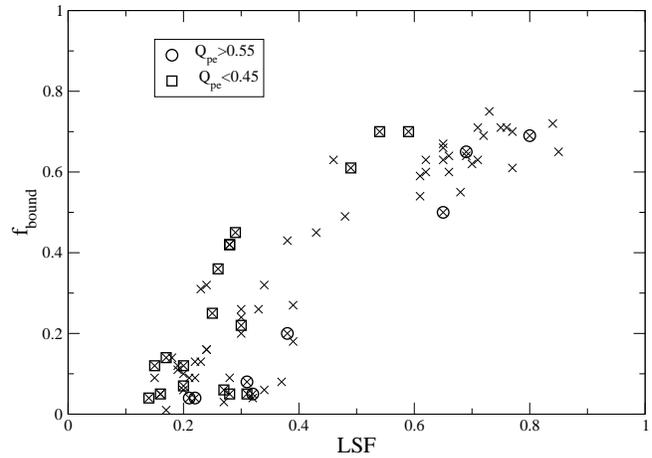}
  \caption{Plot of the influence of the dynamical state of the cluster
    (as measured by $Q_{\rm{pe}}$) at the time of gas removal.  All
    symbols in Figure \ref{mnormLSF} (the standard model simulations)
    are now shown as crosses. Circle and square symbols represent
    simulations where $Q_{\rm{pe}}>0.55$ or $Q_{\rm{pe}}<0.45$
    respectively as indicated in the key.}
\label{mnormLSFQ}
\end{figure}

However, Figure \ref{mnormLSF} shows there is still a considerable 
scatter in the $f_{\rm{bound}}$-$LSF$ trend. Indeed, some clusters
with an $LSF$ of $<0.3$ can retain a significant bound fraction of
stars after gas expulsion (up to 50 per cent).  This is because the 
$LSF$ is not the only parameter influencing stellar mass
loss as a result of gas expulsion. The dynamical state of the cluster,
as measured by $Q_{\rm{pe}}$, is also expected to be important
(\citealp{Goodwin2009ApSS}). Recall from Figure \ref{QfQi} - by the
onset of gas expulsion all clusters had become {\it{approximately}}
virialised. However, even minor deviations from virial equilibrium at
the onset of gas expulsion can significantly effect the cluster
evolution post-gas-expulsion. In Figure \ref{mnormLSFQ}, we again
plot the $f_{\rm{bound}}$-$LSF$ trend, but this time highlighting
simulations whose pre-gas-expulsion virial ratio $Q_{\rm{pe}}$ is
greater than 0.55 (super-virial) or less than 0.45 (sub-virial).
Clusters that are slightly super-virial at the onset of gas expulsion
clearly lie to the lower bounds of $f_{\rm bound}$, whilst those which
are sub-virial are at the higher-end.

\begin{figure}
  \centering \epsfxsize=8.5cm \epsfysize=7cm \epsffile{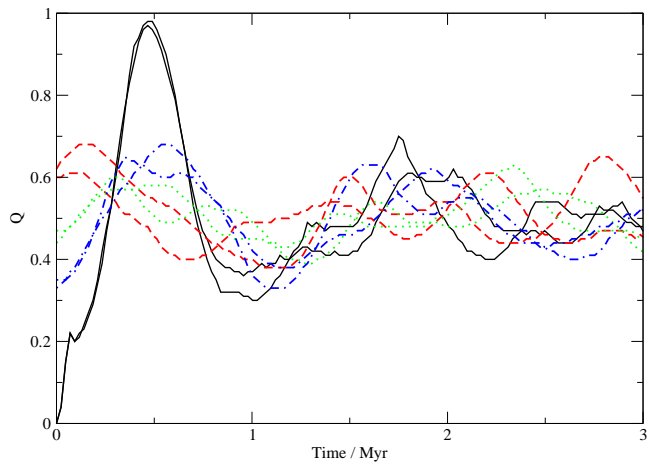}
  \caption{Time evolution of the virial ratio for $Q_{\rm{i}}=0.0$
    (solid, black lines), $Q_{\rm{i}}=0.33$ (dot-dash, blue
    lines), $Q_{\rm{i}}=0.47$ (dotted, green lines), and
    $Q_{\rm{i}}=0.6$ (dashed, red lines). Colour version available online}
\label{Qevol}
\end{figure}

A gravitational system will reach approximate virial equilibrium in 1
-- 2 crossing times, but it will tend to oscillate around exact virial
equilibrium for some time (see also \citealp{Goodwin1997a}).  In 
Figure~\ref{Qevol} we plot the evolution of the virial ratio with
time during the embedded phase. We include representative curves for
simulations with initial virial ratios of $Q_{\rm{i}}=0.0$,
$Q_{\rm{i}}=0.33$, $Q_{\rm{i}}=0.47$ and $Q_{\rm{i}}=0.6$. All curves
are results from simulations with a concentrated gas profile with a
crossing-time of 1.4 Myr. Whilst all clusters reach approximate virial
equilibrium quite quickly they oscillate around $Q=0.5$ for some
time.  And, as shown in fig.~\ref{mnormLSFQ}, the virial ratio they
have at the moment of gas expulsion can make a significant difference
to their ability to form a bound cluster afterwards.

\section{Discussion}

We have shown that initially clumpy and out-of-equilibrium clusters 
will rapidly relax and
erase most of their substructure.  Depending on their initial virial ratios
and substructure this may mean that they may either collapse or 
expand.  This will alter the
importance of the gas potential on the stars meaning that initially
cool clusters are more able to survive gas expulsion, whilst initially
warm clusters are less likely to.  This can remove much of the
dependence of survival after gas expulsion from the importance of the
initial true SFE and place it instead on the
importance of the initial dynamical state of the stars (see 
also \citealp{Verschueren1989}; \citealp{Goodwin2009ApSS}).  We have also shown
that there is a very significant scatter due to both the intrinsic
differences between (statistically the same) clusters (see \citealp{Allison2010}), and on the exact virial ratio at the onset of
gas expulsion.

There is significant observational and theoretical evidence that the
initial distributions of stars are not smooth, nor are they virialised
(see \citealp{Elmegreen2001}; \citealp{Bate2003}; \citealp{Bonnell2003}; \citealp{Bertout2006}; \citealp{Allen2007}; \citealp{Kraus2008}; \citealp{Gutermuth2009}; \citealp{Clarke2010}; \citealp{Bressert2010} and references in all of these papers) which is a natural
consequence of gravoturbulent star formation (see e.g. \citealp{Elmegreen2004};
\citealp{McKee2007}; \citealp{Bergin2007}; \citealp{Clarke2010}).  Indeed,
evidence points towards sub-virial initial conditions for stars (see
\cite{Allison2010} and references therein).
We therefore argue that our initial conditions are far more realistic
than those of a smooth, relaxed cluster as generally used before
(e.g. \citealp{Goodwin1997a}; \citealp{Goodwin2006}; \citealp{Baumgardt2007}).

\cite{Proszkow2009} conduct a large parameter study investigating the key parameters controlling the final bound fraction of clusters that have undergone gas expulsion using smooth and spherical initial stellar distributions. They find a clear trend with SFE although they, too, use sub-virial initial stellar distributions. We note that we also see a trend for increasing bound fraction with increasing SFE (Figure \ref{fboundsfe}), but that it is highly scattered. The source of this scatter is the use of clumpy initial stellar distribution. Therefore the strength of the star formation efficiency as a predictor for the survival of a cluster to mass loss is severely weakened with the use of far more realistic initially clumpy stellar distributions.

It has often previously been assumed that for a cluster to survive it must
have had a high SFE and so the small numbers of
clusters we see must be a high-SFE tail to the SFE distribution  
(see e.g. \citealp{Parmentier2008}).  However, we have shown that
some low-SFE clusters can survive if they are `lucky' enough to have
the right initial conditions.  Indeed, the low survival rates 
found for young clusters of only $\sim
10$ per cent (\citealp{Lada2003}) may be better explained as these
being the few clusters with the `right' initial conditions than being
an extremely high-SFE ($>40$ or $50$ per cent) tail of star formation. 

We note that our simulations are highly idealised. Stars are equal mass, when \cite{Allison2009b} have demonstrated that rapid mass segregation can occur with a more realistic initial mass function.

We do not include any primordial binaries, although their presence can clearly influence the dynamics of a cluster (\citealp{Goodman1989}). The simulations of \cite{Kroupa2001} show that changes in binary fraction, and scattering between stars in the stellar outflow (that can result in binary hardening) can increase the final bound fraction of a cluster. In \cite{Moeckel2010} the binary fraction does not change significantly during the cluster formation process as a result of formation in an initially high stellar density environment - within sub clumps. However their initial conditions are limited to statistics of one, and we have further demonstrated that clumpy initial conditions can result in highly stochastic behaviour.

Furthermore we assume instantaneous gas removal, although \cite{Baumgardt2007} demonstrate that a slower rate of gas removal can result in a higher cluster survival rate.

The length of the embedded phase is fixed at 3 Myr in our simulations although this could vary depending on the nature of the gas removal mechanism. Despite these simplifications, we argue that the idealised nature of the simulations has enabled us to more clearly test the implications of an initially sub-virial, and clumpy stellar distribution. We defer a less idealised study to a later paper.

Our simulations do have an obvious problem, however, in that we use a 
smooth, static background potential for the gas.  This has two main
problems.  Firstly, the initial gas and stellar distribution do not
match, despite the fact that our stars are assumed to have formed from
this background gas.  Secondly, the gas is not able to respond to the
motion of the stars.  (A third, but less important problem, is that we
assume that all of the stars form instantaneously.)

The importance of both problems comes down to how well the motions of
the gas and the stars are coupled.  If conditions are such that both
stars and (at least a significant fraction of the) gas move together
in the potential then we would expect both to collapse or expand
together and the $LSF$ and true star formation efficiency to remain
roughly constant.  However, if the bulk of the gas does not notice the
stars because it is not involved in their formation and the relative
gravitational influence of the stars is small, then our approximations
should be roughly correct.  We would argue that at low true star
formation efficiencies that the bulk of the gas would be uncoupled
from the stars.  We are working on more detailed simulations with a
live background potential which we will present in future papers.

\section{Summary $\&$ Conclusions}
We perform $N$-body simulations of sub-structured, non-equilibrium 
$500 M_\odot$ clusters of $N=1000$ equal-mass stars in a static
background potential.  The mass of gas is varied to simulate star
formation efficiencies (SFEs) of 20 to 40 per cent.  After 
3~Myr of dynamical evolution, the potential is instantaneously 
removed to model the effect of gas expulsion from the cluster.  

Previous work with initially smooth and equilibrium clusters has shown
that there is a critical SFE for the survival of (at least part of)
the cluster of $\sim 30$~per cent (e.g. \cite{Goodwin2006};
\cite{Baumgardt2007} and references therein).  However, it has
also been known that it is the conditions {\it{at the onset}}
of gas expulsion that are important in influencing the evolution of the
star cluster following gas expulsion (\citealp{Verschueren1989},
\citealp{Goodwin2009ApSS}).

Our key results may be summarised as follows.
\begin{enumerate}
\item The true SFE is not a good measure of the survival or otherwise
  of initially out-of-equilibrium clusters.  Even though clusters
  rapidly come into approximate virial equilibrium, their structure
  can change significantly and, unless the gas follows the motions of
  the stars, the ratio of gas mass-to-stellar mass quantified by 
  the local stellar fraction ($LSF$) can change hugely.
\item It is the $LSF$, measured at the onset of gas
  expulsion, that is the key parameter controlling the survival of an
  embedded cluster to gas expulsion (see Figure \ref{mnormLSF}). A
  cluster with a high $LSF$ at the onset of gas expulsion --
  regardless of the initial SFE -- is able to produce a cluster with
  a significant fraction of the initial cluster mass after gas expulsion. 
\item The dynamical state of the cluster, as measured by the virial
  ratio at the onset of gas expulsion, is also
  important. If gas expulsion occurs when the cluster is
  mildly sub-virial (i.e. collapsing), the cluster will suffer lower
  stellar mass loss as a result of gas expulsion. Conversely, a
  cluster that is mildly super-virial (i.e. expanding) at the onset of
  gas expulsion suffers greater stellar mass loss during gas
  expulsion.
\end{enumerate}

The initial SFE of a cluster is almost certainly not a good measure of
the ability of a cluster to survive gas expulsion.  The initial
spacial and kinematic distributions of the stars is crucial, as is how
the stars and gas both evolve once the stars have formed.  It is quite
possible for low-SFE clusters ($< 20$ per cent) to produce bound
clusters after gas expulsion given the right initial conditions.

\section*{Acknowledgements}
MF was financed through FONDECYT grant 1095092, RS was financed through a combination of GEMINI-CONICYT fund 32080008 and a COMITE MIXTO grant, and PA was financed through a CONCICYT PhD Scholarship.

\bibliography{bibfile}

\bsp

\label{lastpage}

\end{document}